\documentclass{aa}
\usepackage[varg]{txfonts}
\usepackage{natbib}
\usepackage[table]{xcolor}
\usepackage{color,array}
\usepackage{arydshln}
\usepackage{xcolor,url}
\usepackage{graphicx}

\bibpunct{(}{)}{;}{a}{}{,}
\makeatletter

\def\zapcolorreset{\let\reset@color\relax\ignorespaces}
\def\colorrows#1{\noalign{\aftergroup\zapcolorreset#1}\ignorespaces}

\begin{document}

        \title{A study on missing lines in the synthetic solar spectrum near the Ca triplet}

        \author{Jessica R. Kitamura\inst{1,3} \and Lucimara P. Martins\inst{1} \and Paula Coelho\inst{2}  }
        
        \offprints{L. Martins, \email{lucimara.martins@cruzeirodosul.edu.br}}

\institute{NAT--Universidade Cruzeiro do Sul, Rua Galv\~ao Bueno, 868, 01506-000, Sao Paulo, SP, Brazil
 \and IAG--Universidade de S\~ao Paulo, Rua do Mat\~ao, 1226, 05508-090, Sao Paulo, SP, Brazil
 \and (ARI--Liverpool John Moores University, IC2 Liverpool Science Park, 146 Brownlow Hill, L3 5RF, Liverpool, United Kingdom)}

\date{Received 03 October 2016 / Accepted 31 January 2017}

\abstract{Synthetic stellar spectra are extensively used for many different applications in
astronomy, from stellar studies (such as in the determination of atmospheric parameters of observed stellar spectra), to extragalactic studies (e.g. as one of the main ingredients of stellar population models). One of the main ingredients of synthetic spectral libraries are the atomic and molecular line lists, which contain the data
required to model all the absorption lines that should appear in these spectra. Although currently available line lists contain millions of lines, a relatively small fraction of these lines have accurate derived or measured transition parameters. As a consequence, many of these lines contain errors in the electronic transition parameters that can reach up to 200\%. Furthermore, even for the Sun, our closest and most studied star,  state-of-the-art synthetic spectra does not reproduce all the observed lines, indicating transitions that are missing in the line lists of the computed 
synthetic spectra. Given the importance and wide range of applications of these models, improvement of their quality is urgently necessary. In this work we catalogued missing lines in the atomic and molecular line lists used for the calculation of the synthetic spectra 
in the region of GAIA, comparing a solar model computed via a recent line list with a high quality solar atlas available in the literature. After that, we attempted the calibration of their atomic parameters with the code ALLiCE; the calibrated line parameters are publicly available for use.}


\maketitle

\section{Introduction}

Synthetic stellar spectra are extensively used in many different applications in astronomy, 
from the determination of atmospheric parameters of observed stellar spectra to the study 
of galaxy integrated spectra (as one of the main ingredients of stellar
population models). 
Generating an accurate synthetic stellar spectrum requires the thermodynamic description of the atmosphere as a function of depth 
\citep[{e.g.}][]{K93,CK04,Px11}, the so-called model atmosphere.   
These model atmospheres may include simplifications for 
complex physical process, such as convection, or computational limitations, 
for example, geometry and/or boundary conditions; often they depend on incomplete 
or imprecise atomic and molecular opacities \citep{Cz15}.

To generate the synthetic stellar spectra, a stellar synthesis code 
\citep[{e.g. SYNTHE;}][]{KA81}
solves the radiative transfer of a given model atmosphere using an atomic and molecular line list. 
Atomic and molecular opacities 
also greatly affect the radiative transfer in stars and consequently their physical 
structure. 
Despite this, 
we are far away from having complete and accurate line lists and, in fact, half of the known lines in the stellar spectra are not 
present in the line lists with wavelengths considered to have good accuracy \citep{K11}.
Besides, detailed spectral models calibrated to a single star (like Sun or Vega) are very 
important but, at the same time, rare and poorly tested \citep[e.g.][]{Cz15}.

In recent decades, there were great efforts to complete and improve the line lists and 
nowadays there are large atomic databases ({\it e.g.} National Institute of Standards and 
Technology-Atomic Spectra Database\footnote{\url{http://www.nist.gov/}} and Vienna Atomic Line 
Database\footnote{\url{http://vald.astro.uu.se}}).

Although the current line lists include millions of absorption lines, 
only a small fraction of these lines were actually accurately measured in laboratory or have accurate parameters derived, and
the uncertainty in the transition parameters for atomic lines can reach up to 200\%. 
Even for the Sun, our closest and most studied star, the synthetic spectrum\ still 
does not completely reproduce all the features and many of the lines are missing in the 
synthetic spectrum \citep{K11}.
For the purpose of spectrophotometric studies, the missing lines have been included
via the so-called ``predicted lines", in which, rather than measured, one or both energy levels of the transition was 
only predicted through quantum mechanics calculations \citep{K92}. These lines are essential to 
better describe the structure of atmospheric models and for spectrophotometric forecasts \citep[e.g.][]{SL96,C14}. 
However, these quantum mechanical predictions are accurate only at a few percent level and the wavelength of these lines 
can be largely incorrect. As such, the inclusion of the predicted lines is inappropriate 
to generate theoretical stellar spectra with high resolution \citep[e.g.][]{B94,CK04,M05}.
As a consequence of this complex scenario, synthetic libraries are only partially able to reproduce observed spectra
and the quality of the model varies with spectral type and wavelength range modelled 
\citep[e.g.][]{bertone05,bertone+08,MC07,C14}.

In an attempt to fill this gap, many groups have been working through the years to improve the quality of atomic data 
used to generate synthetic spectra, focusing on reducing the uncertainty of the transition probabilities 
\citep[e.g.][]{T90,K02,FW06,SS10,B11,P11,Wi11,C12,R13} and the broadening parameters 
\citep[e.g.][]{AO95,BO97,B98,L99,B00,K02,D03,Dm03}. 
Besides, many limitations of these atomic line lists have been approached by different methods and authors,
mainly aimed at chemical analyses of stellar photospheres using high resolution spectra 
\citep[e.g.][]{S97,BW08,B03,J04,H11,SV13,W13}.
However, not only known lines have to be improved, but
the identification of missing lines is necessary to fill significant gaps that are poorly modelled in 
stellar spectra.

In this work, we identify and catalogue missing lines in the atomic and molecular line lists used to generate 
synthetic spectra of the Sun based on a recent line list available in literature, and, when possible, calibrate the atomic lines. 
For this task we used an observed solar spectrum  \citep{W11}, which is an obvious choice since it is the 
highest quality observed stellar spectrum available in the literature.
Additionally, uncertainties involved in the determination of its effective temperature, superficial gravity, 
chemical abundance, etc. are smaller than for any other star. 
The identification of missing lines was performed in the wavelength range from 
$\mathrm {8\,470\,\AA ~\mbox{to}~ 8\,740\,\AA}$ since it is the spectral region covered by 
the Radial-Velocity Spectrometer on the GAIA telescope 
\citep{Lindegren+96,Mignard05} 
launched in 2013. Even though the spectral resolution of the mentioned instrument is relatively low ($\lambda/\Delta\lambda \sim$ 11500 ), we can expect this wavelength region 
to be highly attractive to stellar and galactic studies for the years to come. 
We used the observed solar spectrum available from \citet{W11} and 
compared with the synthetic spectrum generated using the atomic and molecular line list from \citep{S04}, 
updated by \citet{C14}. We use the code ALiCCE 
\citep [Atomic Lines Calibration using the Cross-Entropy Algorithm;][]{M14} to calibrate the atomic parameters.

This paper is structured as follows: in Section 2, we give details about the observed solar spectrum
used in this work; in Section 3 we present the identification and characterisation of the
missing lines in the synthetic solar spectrum; in Section 4 we show the calibration of the atomic
parameters of some of these lines; and, in Section 5, we present our discussion and conclusions.

\section{The observed solar spectrum}

High resolution spectral atlases of the Sun have been produced since the middle of the 20th century \citep[e.g.][]{M40}. 
However, most of the solar spectral atlases are of disk-centre regions. The solar flux spectra, taken over the integrated
disk, are much less common. A flux atlas shows the mean effects of rotation, convection, and centre-to-limb variation. 
Thus, flux spectra are fundamental for comparisons with other stellar spectra and with synthetic spectra \citep{W11}.

The observed solar spectrum used in this work was published by \citet{W11}. These authors
claim that the quality of their spectrum is higher than \citet{K05} because of the more efficient 
subtraction of the telluric lines. The observation was taken at McMath-Pierce Solar Telescope, 
located on Kitt Peak, using the Fourier Transform Spectrograph (hereafter FTS). 
The solar integrated light FTS spectra were obtained mainly on two occasions: 
in 1980--1981 for \citet{K84} atlas and in 1989 for monitoring the irradiance spectrum over the solar cycle \citep{ML91}. 
Both sets were observed near the peaks of sunspot activity. These were the data used by \citet{W11} 
to produce a new flux atlas.

The wavelength coverage of this observed spectrum ranges from $\mathrm {2\,958\,\AA ~\mbox{to}~ 9\,257\,\AA}$ 
and the spectral resolution varies from 350\,000 to 700\,000 ($R=\lambda/\Delta\lambda$). 
The signal to noise in the continuum goes beyond several hundreds. 
In principle, the FTS could have 
observed the entire spectral region in a single integration, however, to reduce the photon noise, six separate observations
were carried out with the spectral coverage in each limited by optical bandpass filters. 
The Doppler shift correction was determined empirically by measuring the solar \ion{Fe}{i} line positions 
and correcting them to the frequencies in \citet{N94}.  Thus, the solar gravitational redshift was also 
removed from the wave number scales.

\citet{W11} made the telluric emission correction, knowing that from $\mathrm {2\,958\,\AA ~\mbox{to}~ 5\,400\,\AA}$ 
the solar spectrum is free from any terrestrial lines. However, weak lines of $\mathrm {H_2O}$ 
begin to appear from $5\,400\,\AA$
 and  $\mathrm {O_2}$ lines also appear from $5\,790\,\AA$. 
The scheme developed by \citet{W11} to correct for the telluric spectrum was to use the solar spectra 
obtained from many sets with different air masses, in the morning or evening, applied to the disk-centre spectra. 
This was carried out because the flux spectra were not taken in suitable airmass sets to allow the transmission
spectra to be extracted. 
They found the best signal to noise and correction with the flux spectrum from October 1989, 
using a disk-centre observation from July 1983. 

\begin{figure*}
                \centering
                \includegraphics[width=19cm]{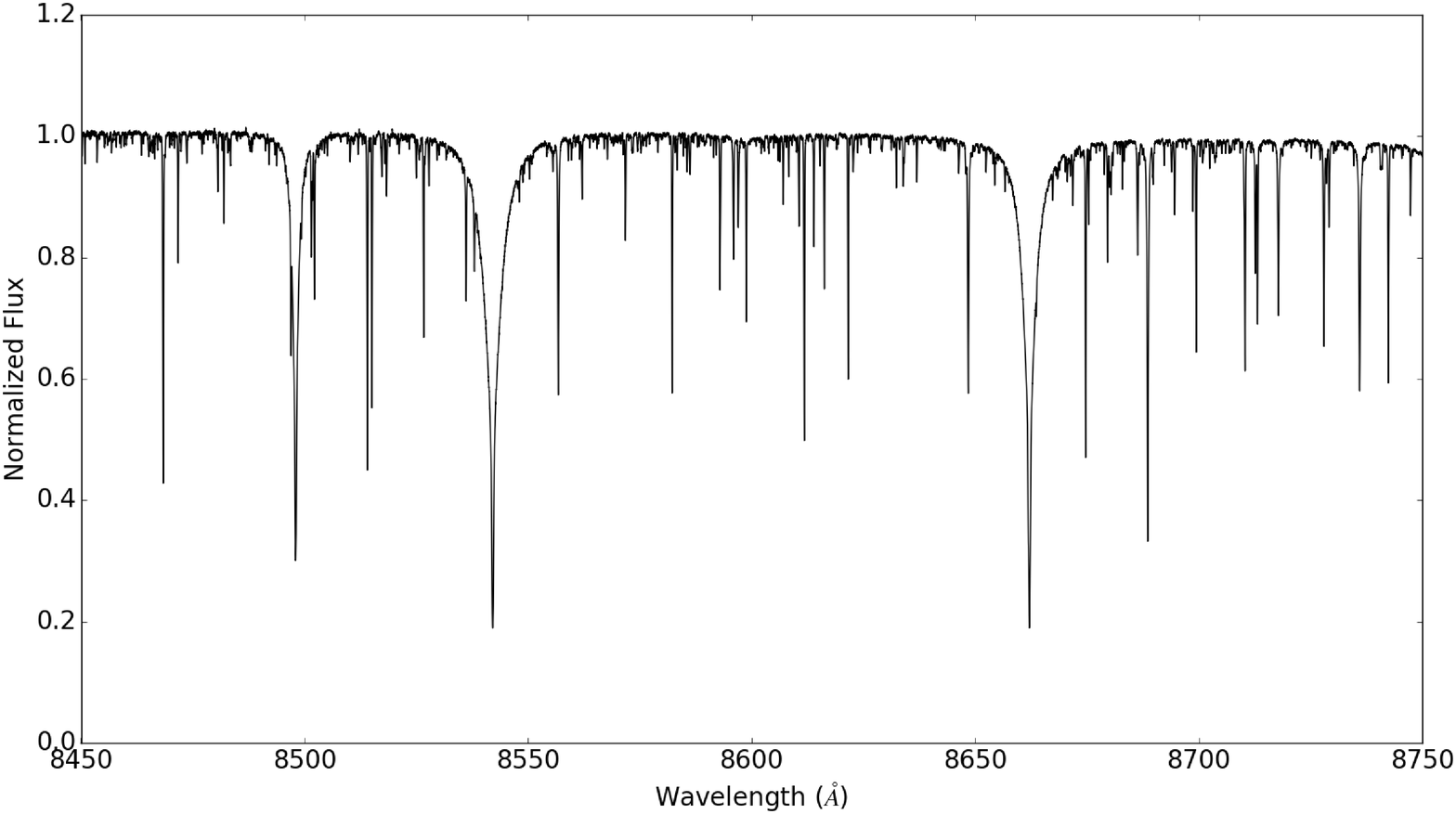}
                \caption{Observed spectrum of the Sun for the wavelength range, 
        from $\mathrm {8\,470\,\AA ~\mbox{to}~ 8\,740\,\AA}$, analysed in this work, as published by \citet{W11}.}
                \label{sunspec}
        \end{figure*}

More details about the observations and the reduction processes are available in \citet{W11}. 
Figure~\ref{sunspec} shows the \citet{W11} spectrum for the spectral region used in this work.

\section{Line identifications}

We used the solar model atmosphere  from \citet{CK04}, 
which is based in ATLAS9 \citep{K70,S04}, to generate the synthetic spectrum of the Sun. There are two different downloadable atmosphere models: 
one using the \citet{AGS05} chemical abundances and another using the \citet{GS98} 
\footnote{\url{http://wwwuser.oats.inaf.it/castelli/sun.html}}. 
Both are tested in this work. The effective temperature of the Sun considered is ${\rm T_{eff}} = $5777~K, 
the surface gravity is $\log g = $4.44 \citep{K70} and convection was taken into account. 
We chose ATLAS9 because it is a static and local thermodynamic equilibrium atmosphere model and is still one of the most commonly used models for chemical abundances studies to generate synthetic stellar libraries, 
and it reproduces well the colours of observed stars \citep{MC07}.

To generate the synthetic spectrum, we used the SYNTHE code \citep{KA81}
in its Linux version published by \citet{S04}. The atomic and molecular line lists used are 
publicly available with ATLAS9 and were updated according to \citet{C14}. 
The spectral synthesis code uses the atomic and molecular line lists to solve the equation of radiative transfer. 
The parameters necessary for the calculation of each absorption line are the central wavelength, 
the energy of the upper and lower levels, the oscillator strength, and the broadening parameters 
(natural, Stark and Van der Waals). The broadening parameters dominate the line wings, 
while the oscillator strengths dominate the line depths. The spectral resolution adopted for this work, 
given by $R=\lambda/\Delta\lambda$, was 676000, 
which is the same resolution of the observed spectrum in this region. To reproduce the line broadening 
from the solar rotational velocity, we applied a rotational 
velocity ($V_{\sin i}$) of 2.4 $km~s^{-1}$ \citep{S78} and 
microturbulence of 1.0 $km~s^{-1}$.

We produced synthetic spectra for three different chemical abundances:
\citet{AGS05} and \citet{GS98}, whose model atmosphere were downloaded from the website of F. Castelli as indicated above, and \citet{A09}. 
To produce a synthetic solar spectrum with the latest abundances we used the same model atmosphere as
with \citet{AGS05} abundances, but changing abundances in SYNTHE code accordingly. 
This approach is valid for small variations in the chemical abundances, which 
should not significantly change the atmosphere structure. 
The three different abundances were tested just to ensure that weak
missing lines were not just an artefact of small abundance changes.
 The comparison between the spectra was carried out visually over the entire spectral 
range studied to identify the missing lines in the 
theoretical spectrum. The telluric spectrum used for the correction of the solar spectrum was also 
inspected to avoid the classification of possible residual subtractions as missing lines.

\begin{figure*}
                \centering
                \includegraphics[width=17cm]{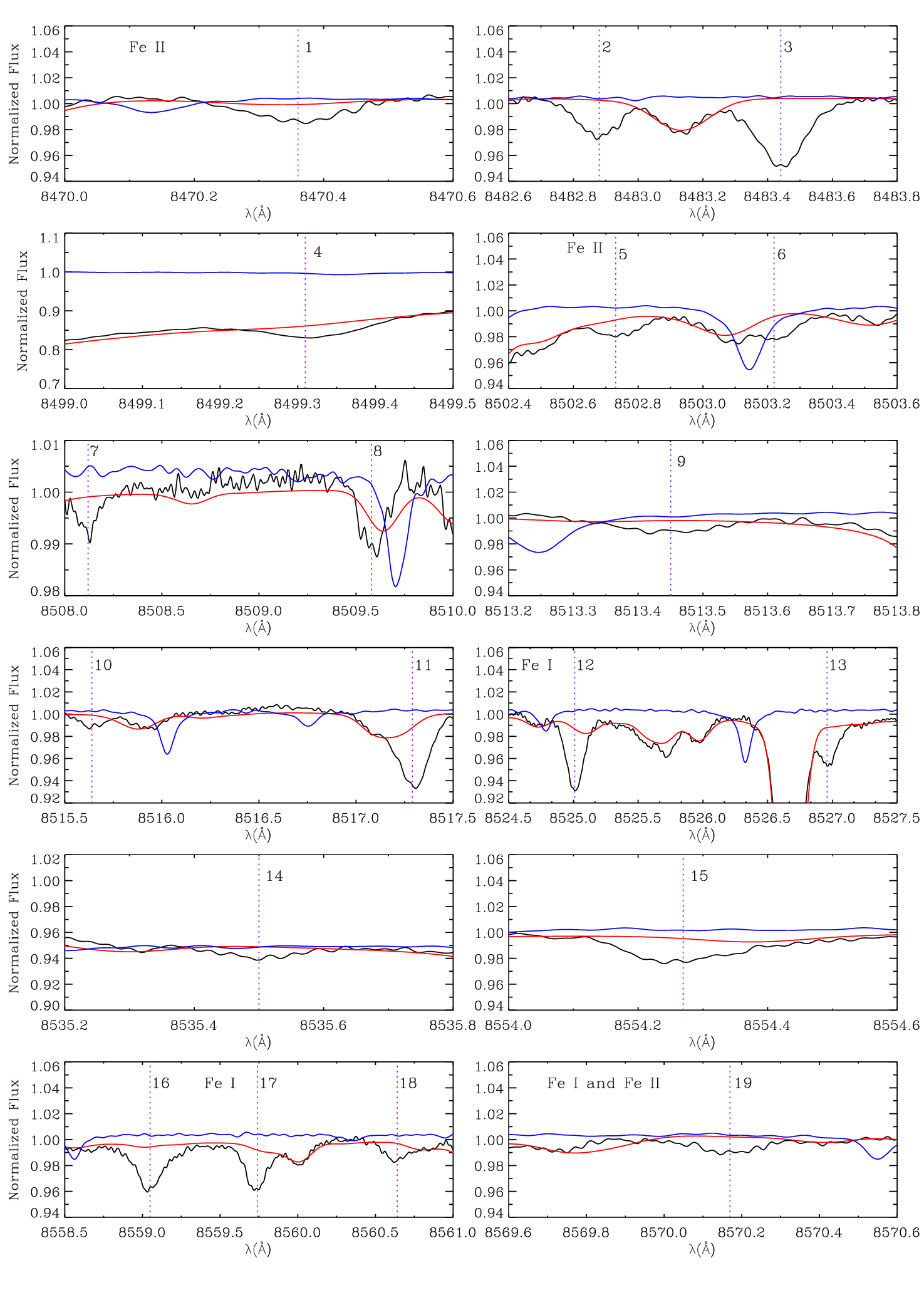}
                \caption{Lines catalogued as absent in the line list (lines 1 to 19 in 
Table~\ref{lineident}). The blue line is the telluric spectrum, the red line is the synthetic spectrum, and the black line is the observed spectrum.}
        \label{lines1}
    \end{figure*}

\begin{figure*}
                \centering
                \includegraphics[width=17cm]{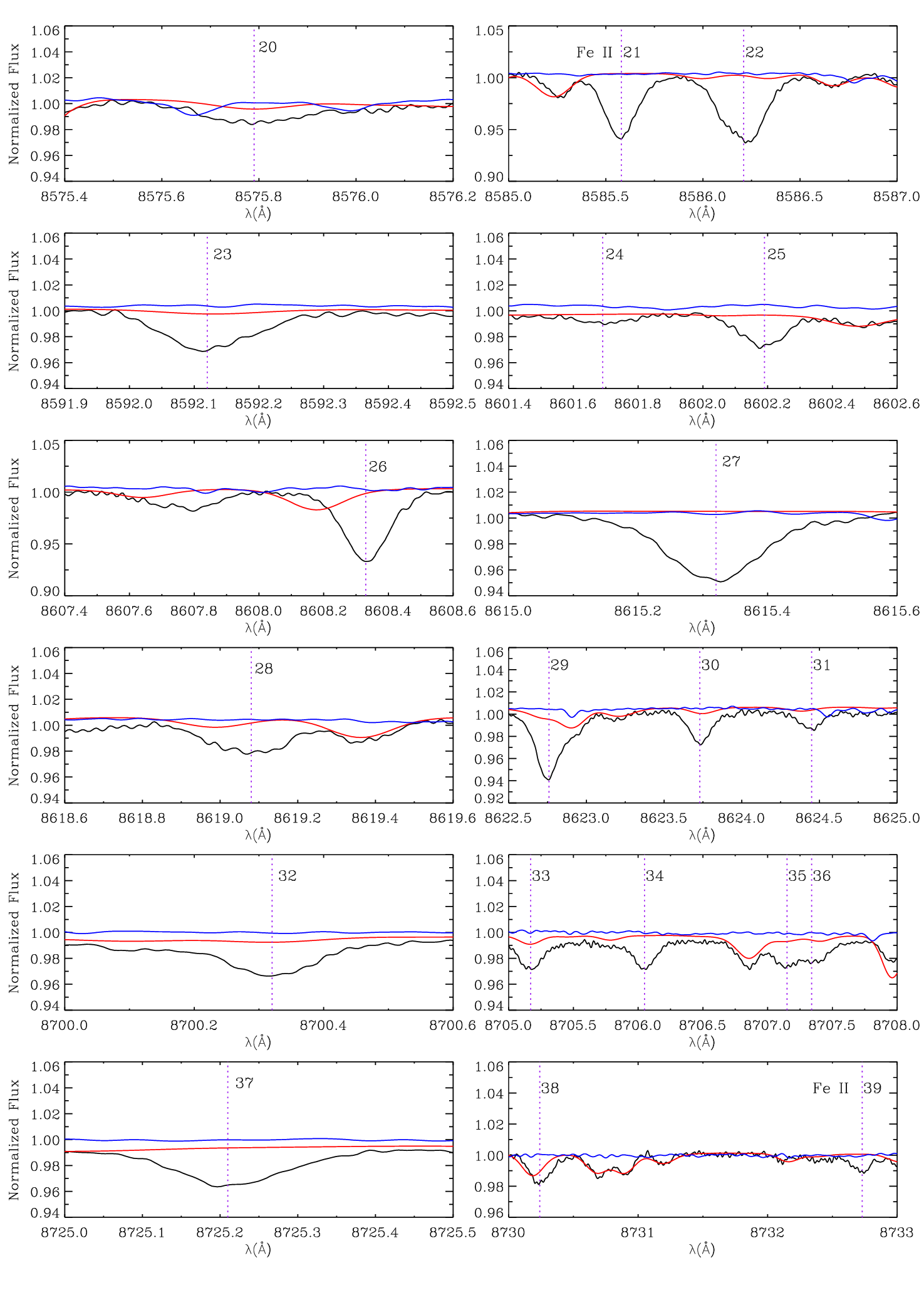}
                \caption{Lines catalogued as absent in the line list (lines 20 to 39 in 
Table~\ref{lineident}). The blue line is the telluric spectrum, the red line is the synthetic spectrum, and the black line is the observed spectrum.}
        \label{lines2}
    \end{figure*}

All lines in this wavelength range that were present in the observed spectrum and absent in the theoretical 
spectrum have been identified and catalogued. We found 39 missing lines in the wavelength range analysed. 
For the characterisation of these lines, each one was fitted by a Gaussian profile. 
\citet{W11} also identified some lines in the spectrum of the Sun and therefore we compared our 
catalogued lines with the lines identified by these authors. Table~\ref{lineident} shows the central wavelength values of the 
catalogued lines, their identification by \citet{W11} (when present), the equivalent width (EW), 
the full width at half maximum (FWHM), and the asymmetry of each line. Figures~\ref{lines1} and ~\ref{lines2} show each of the missing 
lines identified.
        
\begin{table}[h]
        \centering
    \caption{Characterisation of the catalogued lines: Identification by \citet{W11}, central wavelength, EW, FWHM, and asymmetry for each missing line.}
    \begin{footnotesize}
                \begin{tabular}{cp{1cm}ccp{1cm}c}
        \hline
$\#$ & \citeauthor{W11} & $\mathrm{\lambda_{central}}$ (\AA) & EW (\AA) & FWHM $\mathrm{(km~s^{-1})}$ & asymmetry\\
                \hline
        1 & \ion{Cr}{ii} &      8470.36 & 0.0036 & 2.747 & -0.259\\
        2 & \ion{Fe}{i} &       8482.88 & 0.0046 & 2.309 & -0.130\\
        3 & -           &       8483.44 & 0.0078 & 2.379 & -0.274\\
        4 & CN red      &       8499.30 & 0.0179 & 3.311 & 0.003\\
        5 & -           &       8502.73 & 0.0192 & 5.581 & 0.851\\
        6 & CN red      &       8503.22 & 0.0042 & 6.556 & 1.034\\
        7 & -           &       8508.12 & 0.0008 & 1.564 & -0.138\\
        8 & -           &       8509.59 & 0.0025 & 2.431 & 0.173\\
        9 & -           &       8513.45 & 0.0029 & 3.122 & 0.099\\
        10 & -          &       8515.65 & 0.0016 & 2.156 & -0.182\\
        11 & -          &       8517.29 & 0.0130 & 3.040 & 0.651\\
        12 & \ion{Fe}{i} &      8525.01 & 0.0096 & 2.180 & -0.193\\
        13 & -          &       8526.96 & 0.0073 & 2.714 & 0.317\\
        14 & -          &       8535.50 & 0.0011 & 1.977 & -0.047\\
        15 & -          &       8554.27 & 0.0707 & 8.576 & 0.679\\
        16 & \ion{Co}{i} &      8559.05 & 0.0051 & 2.502 & -0.129\\
        17 & \ion{Fe}{i} &      8559.74 & 0.0052 & 2.271 & 0.008\\
        18 & -          &       8560.64 & 0.0079 & 4.084 & 0.579\\
        19 & -          &       8570.17 & 0.0016 & 2.406 & 0.317\\
        20 & CN red     &       8575.79 & 0.0129 & 5.960 & 0.657\\
        21 & \ion{S}{i} &       8585.58 & 0.0106 & 2.642 & -0.228\\
        22 & -          &       8586.21 & 0.0195 & 4.114 & -0.007\\
        23 & \ion{Fe}{i} &      8592.12 & 0.0053 & 2.476 & 0.053\\
        24 & -          &       8601.69 & 0.0093 & 6.157 & 0.530\\
        25 & -          &       8602.19 & 0.0043 & 2.442 & 0.082\\
        26 & -          &       8608.33 & 0.0111 & 2.362 & -0.466\\
        27 & -          &       8615.32 & 0.0080 & 2.335 & -0.468\\
        28 & CN red     &       8619.08 & 0.0075 & 3.992 & 0.254\\
        29 & CN red     &       8622.75 & 0.0075 & 2.242 & -0.129\\
        30 & -          &       8623.73 & 0.0043 & 2.218 & -0.427\\
        31 & -          &       8624.45 & 0.0021 & 2.162 & -0.141\\
        32 & \ion{Fe}{i} &      8700.32 & 0.0039 & 2.254 & -0.860\\
        33 & -          &       8705.17 & 0.0035 & 2.631 & -0.154\\
        34 & -          & 8706.06 & 0.0035 & 2.594 & -0.143\\
        35 & \ion{Mg}{i} & 8707.15 & 1.3182 & 26.072 & 0.827\\
        36 & \ion{Cr}{i} & 8707.34 & 1.7846 & 35.624 & 1.73\\
        37 & -          & 8725.21 & 0.0046 & 2.383 & -0.376\\
        38 & CN red     & 8730.25 & 0.0037 & 2.855 & -0.099\\
        39 & -          & 8732.72 & 0.0013 & 2.281 & 0.099\\
                \hline
                \end{tabular}
        \end{footnotesize}
\label{lineident}
\end{table}

We can see in Table~\ref{lineident} that about half of the lines (21 lines) have widths of about 2.4 $km~s^{-1}$, 
which correspond to the 
net effect of rotational velocity, microturbulence, and macroturbulence in the Sun\footnote{
A significant contribution to the spectral line broadening might come from velocity
fields in the stellar photosphere. In 1D model atmospheres these velocity fields are
represented by microturbulence and macroturbulence, although physically they
have little to do with turbulence \citep{Doyle+14}.}.
We assume that the broadening of these lines is dominated by 
these listed combined effects, rather than an effect intrinsic of the lines. 
We can also speculate that these lines are produced or dominated by one atomic transition as opposed to molecular
transitions or a blend of lines. 
Another reason that these lines are produced or dominated by only 
one electronic transition is their symmetry. Asymmetric lines are more likely produced by  a blend of lines, whether atomic or molecular.  
The asymmetry was measured through the skewness of the line, where values
close to zero mean very symmetrical distributions.
Thirteen of these lines have asymmetry values close to zero 
(lines 2, 8, 10, 12, 16, 17, 23, 25, 29, 31, 33, 34, and 39). For these lines, |asymmetry| <0.20. 
However, line 29 was identified by \citet{W11} as due to a CN transition, so we cannot rule out that
some of these lines come from molecular transitions. 
Another caveat that might be taken into account is that
asymmetries and wavelength shifts in stars might be correlated with 
convection in the sense that warm, rising convective elements are blueshifted and cool and falling convective elements are redshifted. However, according to \citet{CONFIRMARdravins98} 
this effect in the Sun is around 300 m/s, which is much smaller than visible asymmetries
seen here. For the value of the central wavelength, this implies in a $\Delta \lambda$ of 0.0085 \AA\ 
at 8500 \AA,
which is also minor for the values observed here.
Signatures with much larger widths are more likely to be produced by a blend of lines,
although single atomic lines can be broadened by hyperfine and isotopic splitting.
\citep{Wahlgren05}.
Two signatures have widths that are smaller than 2.0 $km~s^{-1}$ (lines 7 and 14). 
This is possibly due to measurement errors caused by the difficulty of adjusting these signatures, 
which are in regions with a high density of lines.

After characterising each signature, we looked for the atomic data of these missing lines in the 
NIST database to verify whether their absence was not only a matter of outdated line lists. 
The atomic data collected from NIST were measured in the laboratory. 
Table~\ref{linenist} shows the relation between catalogued missing lines and the data found in NIST.

\begin{table*}[h]
        \centering
    \caption{Catalogued lines and their relative NIST atomic parameters.}
    \begin{footnotesize}
                \begin{tabular}{|p{0.4cm}||c|c|c|c|c|c|c|c|}
        \hline
$\#$ & \bf $\mathrm{\lambda_{ident}}$ (\AA) & \bf $\mathrm{\lambda_{NIST}}$ (\AA) & \bf Element & \bf log {\it gf} & \bf $\mathrm{E_i (cm^{-1})}$ & \bf $\mathrm{E_k (cm^{-1})}$ & \bf Element Code & \bf Quality \\
                \hline
   1 & 8470.36 & 8470.37  & \ion{Fe}{ii} & -2.5 & 54275.649 & 66078.272 & 26.01 & D \\
                \hline
   2 & 8482.88 & 8482.67  & \ion{Ho}{i}  & -    & -         & -         & 67.00 & - \\
   2 & 8482.88 & 8482.684 & \ion{Sc}{ii} & -    & 66048.39  & 77833.88  & 21.01 & - \\
                \hline
   3 & 8483.44 & 8483.39  & \ion{Mo}{i}  & -    & -         & -         & 42.00 & - \\
   3 & 8483.44 & 8483.56  & \ion{Ru}{i}  & -    & 32391.95  & 44176.23  & 44.00 & - \\
                \hline
   4 & 8499.30 &8499.34\tablefootmark{a}   & \ion{Fe}{i}  &  -   & -         & -         & 26.00 & - \\
                \hline
   5 & 8502.73 & 8502.7   & \ion{Tb}{i}  & -    & -         & -          & 65.00 & - \\
   5 & 8502.73 & 8502.714 & \ion{Fe}{ii} & -    & 92116.529 & 103874.261 & 26.01 & - \\
   5 & 8502.73 & 8502.714 & \ion{Fe}{ii} & -    & 93328.553 & 105086.265 & 26.01 & - \\
                \hline
   6 & 8503.22 & 8503.212 & \ion{Co}{ii} & -0.14& 97062.844 & 108819.860 & 27.01 & C+ \\
                \hline
   7 & 8508.12 & 8508.08  & \ion{Lu}{i} &       -   & -         & -          & 71.00 & - \\
                \hline
   8 & 8509.59 & 8509.60\tablefootmark{a}  & \ion{Fe}{i}  &  -   & -         & -         & 26.00 & -   \\
                \hline
   9 & 8513.45 & 8513.38 & \ion{Br}{i}   & -     & -         & -          & 35.00 & - \\
   9 & 8513.45 & 8513.5  & \ion{Hg}{i}   & -     & 71336.005 & 83078.8    & 80.00 & - \\
   9 & 8513.45 & 8513.57 & \ion{La}{i}   & -     & -         & -          & 57.00 & - \\
                \hline
   10 & 8515.65 & 8515.475 & \ion{S}{ii}  & -1.341 & 121528.72 & 133268.68 & 16.01 & C \\
                \hline
   11 & 8517.29 & 8517.37  & \ion{Li}{ii} & -     & -         & -         & 3.01 & - \\
                \hline
   12 & 8525.01 & 8525.029 & \ion{Fe}{i}  & -     & 36975.588 & 48702.535 & 26.00 & - \\
                \hline
   13 & 8526.96 & 8526.99 & \ion{Nb}{i}   & -     & -         & -         & 41.00 & D \\
   13 & 8526.96 & 8527.03 & \ion{Ti}{iii} & 0.08  & 169615.12 & 181339.27 & 22.02 & - \\
                \hline
   14 & 8535.50 & \multicolumn{7}{l|}{\it No lines are available in NIST with this wavelength} \\
                \hline
   15 & 8554.27 & \multicolumn{7}{l|}{\it No lines are available in NIST with this wavelength} \\
        \hline
   16 & 8559.05 & \multicolumn{7}{l|}{\it No lines are available in NIST with this wavelength} \\
                \hline
   17 & 8559.74 & 8559.741 & \ion{Fe}{i} & -       & 41178.412 & 52857.804 & 26.00 & - \\
                \hline
   18 & 8560.64 & 8560.54  & \ion{Nb}{i} & -       & -         & -         & 41.00 & - \\
                \hline
   19 & 8570.17 & 8570.099  & \ion{Fe}{ii} & -     & 92358.625 & 104023.921 & 26.01 & - \\
   19 & 8570.17 & 8570.099  & \ion{Fe}{i}  & -     & 45509.152 & 57174.430  & 26.00 & - \\
                \hline
   20 & 8575.79 & 8575.78   & \ion{Te}{ii} & -     & -         & -          & 52.01 & - \\
   20 & 8575.79 & 8575.87   & \ion{Nb}{i}  & -     & -         & -          & 41.00 & - \\
   20 & 8575.79 & 8575.92   & \ion{Ta}{i}  & -     & -         & -          & 73.00 & - \\
                \hline
   21 & 8585.58 &  8585.4403 & \ion{Fe}{ii} & -0.47 & 90386.533 & 102030.965 & 26.01 & C+ \\
   21 & 8585.58 &  8585.52   & \ion{Sr}{iii}& -     & 269388.34 & 281032.70  & 38.02 & - \\
                \hline
   22 & 8586.21 & \multicolumn{7}{|l|}{\it No lines are available in NIST with this wavelength} \\
                \hline
   23 & 8592.12 & 8592.22   & \ion{In}{ii} & -1.59  & 135999.37 & 147634.61  & 49.01 & D+  \\
                \hline
   24 & 8601.69 & \multicolumn{7}{l|}{\it No lines are available in NIST with this wavelength} \\
        \hline
   25 & 8602.19 & \multicolumn{7}{l|}{\it No lines are available in NIST with this wavelength}\\
                \hline
   26 & 8608.33 & 8608.3067   & \ion{Cs}{ii} & -     & 157572.1025 & 169185.5968 & 55.01 & - \\
   26 & 8608.33 & 8608.3882   & \ion{Cs}{ii} & -     & 157572.1025 & 169185.4867 & 55.01 & - \\
                \hline
   27 & 8615.32 & \multicolumn{7}{l|}{\it No lines are available in NIST with this wavelength} \\
        \hline
   28 & 8619.08 & \multicolumn{7}{|l|}{\it No lines are available in NIST with this wavelength} \\
        \hline
   29 & 8622.75 & \multicolumn{7}{|l|}{\it No lines are available in NIST with this wavelength} \\
                \hline
   30 & 8623.73 & 8623.805    & \ion{Ar}{ii} & - & 189437.7396 & 201030.3698 & 18.01 & - \\
                \hline
   31 & 8624.45 & \multicolumn{7}{|l|}{\it No lines are available in NIST with this wavelength} \\
                \hline
   32 & 8700.32 & 8700.25   & \ion{In}{i} & - & 32915.539 & 44406.31 & 49.00 & - \\
                \hline
   33 & 8705.17 & \multicolumn{7}{l|}{\it No lines are available in NIST with this wavelength} \\
        \hline
   34 & 8706.06 & \multicolumn{7}{|l|}{\it No lines are available in NIST with this wavelength} \\
                \hline
   35 & 8707.15 & 8707.048 & \ion{O}{ii} & - & 246483.317 & 257965.11 & 8.01  & - \\
   35 & 8707.15 & 8707.14  & \ion{Mg}{i} & - & 49346.729  & 60828.41  & 12.00 & - \\
   35 & 8707.15 & 8707.215 & \ion{Tc}{i} & - & 32620.38   & 44101.99  & 43.00 & - \\
                \hline
   36 & 8707.34 & \multicolumn{7}{l|}{\it No lines are available in NIST with this wavelength} \\
        \hline
   37 & 8725.21 & \multicolumn{7}{|l|}{\it No lines are available in NIST with this wavelength} \\
        \hline
   38 & 8730.24 & \multicolumn{7}{|l|}{\it No lines are available in NIST with this wavelength} \\
                \hline
   39 & 8732.75 & 8732.679   & \ion{Sr}{iii} & -    & 270011.52 & 281459.61  & 38.02 & - \\
   39 & 8732.72 & 8732.746   & \ion{Fe}{ii}  & 0.02 & 98568.907 & 110016.915 & 26.01 & E \\
                \hline
                \end{tabular}
        \end{footnotesize}
\tablefoot{
\tablefoottext{a}{Private communication from Gillian Nave (NIST)}}
\label{linenist}
\end{table*}

In Table~\ref{linenist}, the column $\lambda_{Ident}$ concerns all the lines identified as missing from the comparison 
of the observed solar spectrum with the synthetic spectrum listed in Table~\ref{lineident}.
The data in columns $\lambda_{NIST}$, Element, log {\it gf}, and $E_i$ and $E_k$ energies are the values from the
NIST database that correspond to the identified wavelengths. 
The data in column \emph{Element Code} is the code used by Kurucz to characterise each ion. 
From all the identified missing lines, NIST suggests that about a quarter are from Fe transitions. 
For some of the identified lines, we have two or 
more matching lines found in NIST. This is likely because we allowed a search inside a range 
of $0.2\,\AA$ from the measured central wavelength 
to account for possible errors in the central wavelength measurement.

All lines found in NIST with measured log {\it gf} were included in the atomic line list used in the 
spectral synthesis code. Unfortunately, there are very few cases where this happens (seven lines) and we did 
not observe any significant improvement in the synthetic spectra generated after this inclusion. 
This means that despite the absence of these lines in the atomic list used,
the missing lines are due to transitions other than those catalogued in NIST.

In an attempt to find candidates for the missing lines in the line list, 
we searched for them in the VALD database. The values from VALD are not necessarily measured in laboratory
and many of them are theoretically calculated or empirically calibrated.

From the 39 lines missing in the solar spectrum characterised in this work, we found counterparts for 22 on
NIST, although only 7 have a measured log {\it gf}, as previously mentioned. On VALD we found counterparts for 
16 of these 22 lines, although 9 of them were already included in the Kurucz line list (lines 1, 7, 8, 11, 13, 18, 20, 30, and 35). 
The candidates found in VALD are listed in Table~\ref{valdlines}. 
Lines indicated in grey in this table are already present on the line list, either with the
same exact parameters or with values very close to those found. 
For some
of these lines we found more than one counterpart in VALD, which were not found in NIST. 

All lines from Table~\ref{valdlines} that are missing in the line list
are lines from Fe ions, 
which is not surprising given that Fe has a complex electron configuration and is very abundant. 
Weak lines from Fe are important 
because they are frequently used for the measurement of chemical abundances of stars. 
Since the atomic parameters from VALD are not obtained from laboratory, we decided to calibrate these iron lines before including them in the line list.

\begin{table*}
        \centering
    \caption{Fe lines and their relative VALD atomic parameters.}
    \begin{footnotesize}
        \begin{tabular}{c|c|c|c|c|c|c|c|c|c}
                \hline
  $\#$ &  \bf $\mathrm{\lambda_{ident}}$ (\AA) & \bf $\mathrm{\lambda_{VALD}}$ (\AA) & \bf log {\it gf} & \bf Elem. & \bf $\mathrm{E_i~(cm^{-1})}$ & \bf $\mathrm{E_k~(cm^{-1})}$ & \bf log $\Gamma_{R}$ & \bf log $\Gamma_{S/N_{e}}$ & \bf log $\Gamma_{W/N_{H}}$ \\
            \hline
            \colorrows{\color{gray}}
1& 8470.36 & 8470.359\tablefootmark{b}& -2.519 & \ion{Fe}{ii} & 54275.6400 & 66078.2700 & 0.000 & 0.000 & 0.000 \\
1& 8470.36 & 8470.365\tablefootmark{d}& -2.734 & \ion{Fe}{ii} & 54275.6490 & 66078.2720 & 8.560 & -6.530 & -7.900 \\
1& 8470.36 & 8470.390\tablefootmark{a}& -1.443 & \ion{Fe}{i} & 47177.2340 & 58979.8230 & 8.47 & -4.41 & -7.45 \\
                \hline
                \colorrows{\color{black}}
4 & 8499.34 & 8499.330\tablefootmark{a}& -0.693 & \ion{Fe}{i} & 47017.1880 & 58779.5900 & 8.470 & -4.560 & -7.430 \\
                \hline
5 & 8502.70 & 8502.670\tablefootmark{a}& -6.818 & \ion{Fe}{i} & 8154.7140 & 19912.4950 & 3.520 & -6.280 & -7.850 \\
5 & 8502.70 & 8502.705\tablefootmark{d}& -3.059 & \ion{Fe}{ii} & 92116.5290 & 103874.2610 & 8.920 & -5.840 & -7.750 \\
5 & 8502.70 & 8502.720\tablefootmark{d}& -5.427 & \ion{Fe}{ii} & 93328.5530 & 105086.2650 & 8.770 & -5.700 & -7.500 \\
                \hline
                \colorrows{\color{gray}}
7 & 8508.12 & 8508.112\tablefootmark{e}& -0.040 & \ion{Lu}{i} & 23524.2400 & 35274.5000 & 0.00 &  0.00 &  0.00 \\
       \hline
8 & 8509.60 & 8509.617\tablefootmark{a}& -3.436 & \ion{Fe}{i} & 35257.3240 & 47005.5060 & 8.240 & -5.330 & -7.550 \\
       \hline
11 & 8517.29 & 8517.369\tablefootmark{f}& -0.672 & \ion{Li}{ii} & 490071.1000 & 501808.5900 & 10.410 & -5.570 & 0.000 \\
       \hline
       \colorrows{\color{black}}
12 & 8525.01 & 8525.011\tablefootmark{a}& -1.523 & \ion{Fe}{i} & 46889.1420 & 58616.1100 & 8.210 & -4.140 & -7.380 \\
12 & 8525.01 & 8525.026\tablefootmark{a}& -3.599 & \ion{Fe}{i} & 36975.5880 & 48702.5350 & 7.730 & -5.990 & -7.800\\
12 & 8525.01 & 8524.999\tablefootmark{d}& -7.819 & \ion{Fe}{ii} & 86124.3480 & 97851.3320 & 8.960 & -5.680 & -7.710 \\
12 & 8525.01 & 8525.060\tablefootmark{d}& -8.023 & \ion{Fe}{ii} & 113056.8470 & 124783.7480 & 8.610 & -4.370 & -7.420 \\
                \hline
                \colorrows{\color{gray}}
13 & 8526.96 & 8526.958\tablefootmark{g}& -1.560 & \ion{Nb}{i} & 10922.7400 & 22647.0300 & 0.00 & 0.00 & 0.00 \\
13 & 8526.96 & 8527.060\tablefootmark{i}& 0.138 & \ion{Ti}{iii} & 169615.1200 & 181339.2700 & 9.090 & -4.950 & -7.520 \\
                \hline
                \colorrows{\color{black}}
17 & 8559.74 & 8559.738\tablefootmark{a}& -1.510 & \ion{Fe}{i} & 41178.4120 & 52857.8040 & 8.190 & -5.710 & -7.710 \\
                \hline \colorrows{\color{gray}}
18 & 8560.64 & 8560.538\tablefootmark{g}& -1.690 & \ion{Nb}{i} & 8705.3200 & 20383.6200 & 0.00 & 0.00 & 0.00 \\
                \hline
                \colorrows{\color{black}}
19 & 8570.17 & 8570.094\tablefootmark{a}& -3.231 & \ion{Fe}{i} & 45509.1520 & 57174.4300 & 8.010 & -3.990 & -7.260 \\
19 & 8570.17 & 8570.081\tablefootmark{d}& -2.358 & \ion{Fe}{ii} & 92358.6250 & 104023.9210 & 9.010 & -5.900 & -7.740 \\
                \hline
                \colorrows{\color{gray}}
20 & 8575.79 & 8575.879\tablefootmark{g}& -1.710 & \ion{Nb}{i} & 12357.7000 & 24015.1100 & 0.00 & 0.00 & 0.00 \\
20 & 8575.79 & 8575.901\tablefootmark{j}& -2.060 & \ion{Ta}{i} & 13351.4500 & 25008.8300 & 0.00 & 0.00 & 0.00 \\
                \hline          
21 & 8585.58 & 8585.437\tablefootmark{d}& -1.011 & \ion{Fe}{ii} & 90386.5330 & 102030.9650 & 8.520 & -5.180 & -7.480 \\
\colorrows{\color{black}}
21 & 8585.58 & 8585.540\tablefootmark{d}& -5.930 & \ion{Fe}{ii} & 102952.1700 & 114596.4620 & 8.790 & -5.020 & -7.360 \\
                \hline
                \colorrows{\color{gray}}
30 & 8623.73 & 8623.850\tablefootmark{l}& -0.780 & \ion{Ar}{ii} & 189437.7360 & 201030.3000 & 0.000 & 0.000 & 0.000 \\
                \hline
35 & 8707.15 & 8707.135\tablefootmark{l}& -2.760 & \ion{Mg}{i} & 49346.7290 & 60828.4100 & 0.000 & 0.000 & 0.000 \\
                \hline
39 & 8732.73 & 8732.634\tablefootmark{d}& -3.392 & \ion{Fe}{ii} & 93840.4050 & 105228.5590 & 8.750 & -5.650 & -7.500 \\
39 & 8732.73 & 8732.746\tablefootmark{d}& -0.080 & \ion{Fe}{ii} & 98568.9070 & 110016.9150 & 8.920 & -5.730 & -7.690 \\
39 & 8732.73 & 8732.823\tablefootmark{d}& -1.424 & \ion{Fe}{ii} & 100492.0250 & 111939.9320 & 9.040 & -5.740 & -7.680 \\
\colorrows{\color{black}}
39 & 8732.73 & 8732.852\tablefootmark{a}& -6.238 & \ion{Fe}{i} & 45913.4970 & 57361.3660 & 8.030 & -4.890 & -7.290 \\

                \hline
                \colorrows{\color{black}}
        \end{tabular}
    \label{valdlines}
        \end{footnotesize}
        \tablefoot{log $\Gamma_{R}$ is the value of radiation pressure, log $\Gamma_{S/N_{e}}$ is the value of the Stark broadening, and log $\Gamma_{W/N_{H}}$ is the value for the Van der Waals broadening.\\
                \tablefoottext{a}{\citep{RK14}\footnote{http://kurucz.harvard.edu/linelists/gfnew/}}
                \tablefoottext{b}{\citep{RU}}
                \tablefoottext{c}{\citep{BAJ}}
                \tablefoottext{d}{\citep{K13}}
                \tablefoottext{e}{\citep{WV}}
                \tablefoottext{f}{\citep{WSG}}
                \tablefoottext{g}{\citep{DLa}}
                \tablefoottext{h}{\citep{MFW}}
                \tablefoottext{i}{\citep{K10}}
                \tablefoottext{j}{\citep{CBcor}}
                \tablefoottext{l}{\citep{KP}}
}
\end{table*}

\section{Line calibration}

We used the code ALiCCE 
\citep[Atomic Lines Calibration using the Cross-Entropy Algorithm;][]{M14} for the calibration of the atomic parameters.
The ALiCCE code was developed to automatically calibrate atomic lines using the
cross-entropy method \citep[e.g.][]{R97,R99,M04,K06,B05}. The cross-entropy
method is a general Monte Carlo approach to combinatorial and
continuous multi-extremal optimisation and importance sampling,
which is a general technique for estimating properties of a particular
distribution using samples generated randomly
from a different statistical distribution rather than the distribution of interest 
\citep{R97,R99,kroese+06}.

For each iteration, ALiCCE generates N different atomic line lists with
each atomic parameter to be calibrated varying inside a given interval. It then 
makes external calls to the spectral synthesis code SYNTHE for each of the N lists.
The output spectra generated is then compared with the observed spectrum of the Sun 
(the comparison star used for this work) and a performance function is calculated 
for each of the N lists, which is a measure of how well the synthetic spectrum
represents the observed spectrum. The lists are then ranked and the interval for each atomic
parameter is recalculated based on the mean and standard deviation of the
top 5\% solutions. The process starts again until the stopping criteria is fulfilled.
More details about the code can be found in \citet{M14}.

In this work we calibrated only the oscillator strength but not the broadening values since 
the width of the lines are dominated by the rotational velocity
 (plus micro and macro turbulence) of the Sun.  
All possible atomic transitions for a given missing line, as shown in Table~\ref{valdlines},
were calibrated together. However, since the wavelength range between each missing line 
was large enough, we calibrated them separately, running ALiCCE for each missing line
identified.  
The log {\it gf} values obtained from VALD were used as initial guesses. 
Table~\ref{resultallice} shows the values obtained for each line calibrated by ALiCCE.

\begin{table}[h]
    \centering
    \caption{Comparison between VALD and ALiCCE atomic parameters.}
        \begin{tabular}{|c|c|c|c|}
                \hline
   & &VALD & ALiCCE\\
   $\#$ & $\lambda$ (\AA) & log {\it gf} & log {\it gf}\\
 \hline
  4 & 8499.330 & -0.693 & -1.066 $\pm$ 0.001\\ 
 \hdashline
 5 & 8502.705 & -3.059 & no convergence\\
 \hdashline
 12 & 8524.999 & -7.819 & no convergence\\
 12 & 8525.011 & -1.523 & -1.223$\pm$ 0.001\\
 12 & 8525.027 & -2.208 & -2.208 $\pm$ 0.001\\          
 12 & 8525.060 & -8.023 & no convergence\\
  \hdashline
 17 & 8559.738 & -1.509 & -1.941 $\pm$ 0.004\\
  \hdashline
 19 & 8570.081 & -2.385 & 0.389 $\pm$ 0.002\\
 19 & 8570.095 & -3.281 & -1.751 $\pm$ 0.014\tablefootmark{1}\\
  \hdashline
 21 & 8585.540 & -5.93 & no convergence\\
  \hdashline
 39 & 8732.852 & -6.238 & no convergence\\

                \hline
        \end{tabular}
        \tablefoot{\tablefoottext{1}{The  8570.20 line was not included in the calibration because it is very similar to 8570.095 line. We believe they are the same line.}}
\label{resultallice}
\end{table}

\begin{figure*}[h]
                \centering
                \includegraphics[width=17cm]{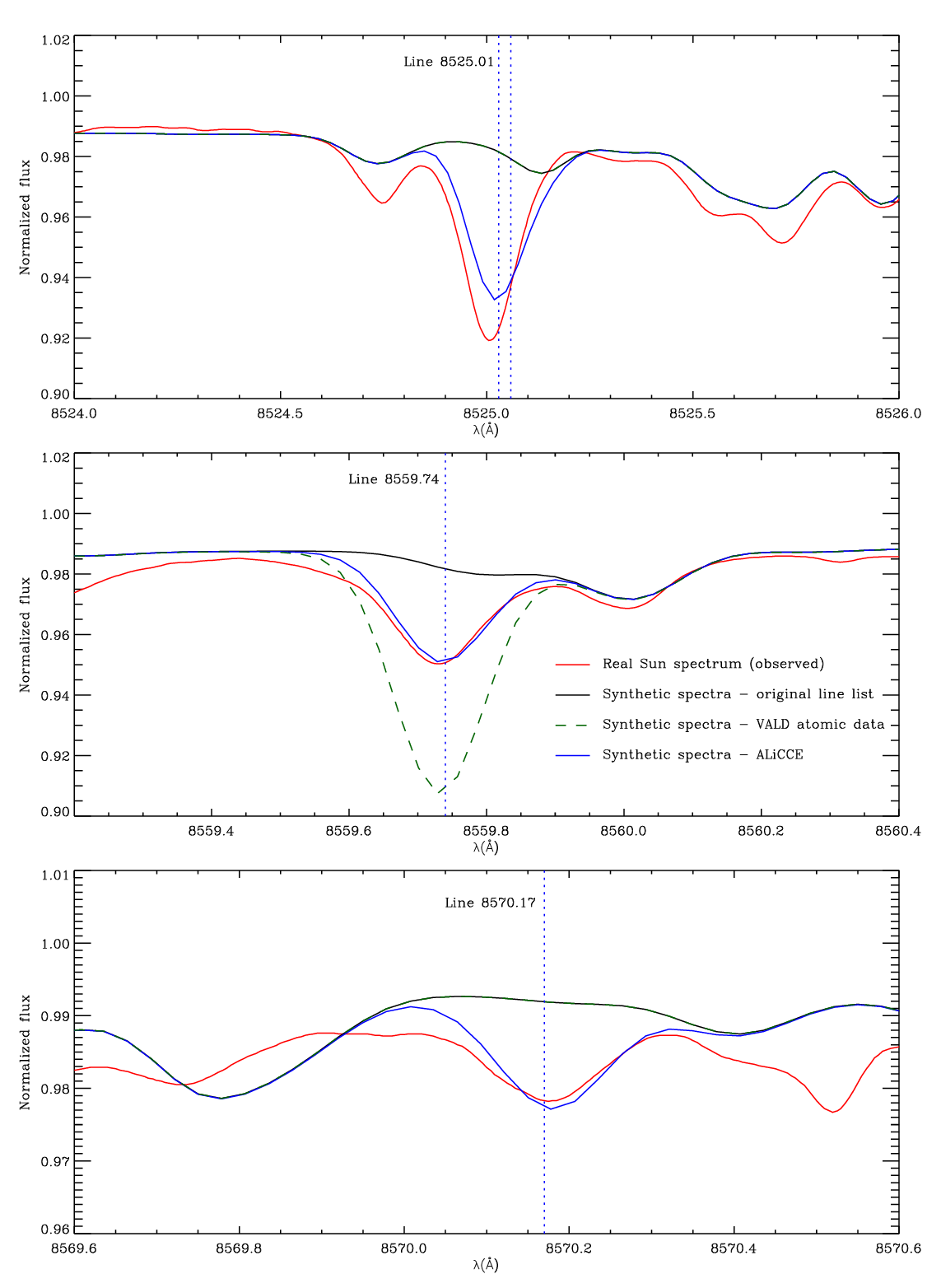}
                \caption{Comparison between the observed spectrum (red line), 
      the synthetic spectrum without the identified and catalogued missing lines (dashed green line), the synthetic spectrum with the lines included in the line list and with the atomic parameters found at VALD (black line), and the synthetic spectrum with the lines calibrated by ALiCCE (blue line).}
      \label{calib_lines}
        \end{figure*}

A significant improvement in the reproduction of the 
identified lines was obtained after the calibration of the log {\it gf} with ALiCCE. These results can be 
seen in Figure~\ref{calib_lines}, which show the comparison of synthetic spectra generated without the presence 
of the lines, values of the atomic parameters from VALD, and log {\it gf} values obtained by ALiCCE. 

The oscillator strengths could not be derived for five lines given that they did not converge in less than 
500 iterations and had errors greater than 0.010 (see details of error evaluation in \citealt{M14}).

\section{Discussion and conclusion}

Improving the quality of the synthetic stellar spectra is of paramount importance
to our understanding of stars and galaxies. Synthetic spectra libraries has an advantage over 
empirical spectra ones, which is the possibility of generating spectra with any atmospheric
parameters desired for any chemical abundance pattern desired. 
However, synthetic
spectra also have limitations, since they can only be as good as the ingredients used to generate
them.  

One of the major problems faced by theoretical stellar spectra is the 
uncertainty and incompleteness 
of the atomic and molecular line lists used to generate them. 
There are still many lines present in stellar spectra that are unknown. 
Moreover, among the millions of known lines present in the line lists, few of 
them have accurate and precise values, whether they are measured or computed.

We have identified and catalogued missing lines in the atomic and molecular 
line lists 
used for producing synthetic stellar spectra in the spectral 
region of $\mathrm {8\,470\,\AA ~\mbox{to}~ 8\,740\,\AA}$.
The line lists that we used are based on previous work by \citet{K70, KA81, CK04} 
and updated according to \citet{C14}.
For this, we used the observed spectrum of the Sun published by \citet{W11} and we generated the 
synthetic spectrum using the spectral synthesis code SYNTHE \citep{KA81} with the model atmosphere of the Sun 
generated by ATLAS9 \citep{K70,S04}. 

We found 39 lines missing from the atomic and molecular line lists within the analysed wavelength region. 
We performed a characterisation of each line by measuring their equivalent widths and their widths at half maximum (FWHM). 
From these values and from the analysis of the symmetry of the lines, we conclude 
that about one-third ($\sim$ 36\%) of the identified lines can be produced or are dominated by a single atomic transition. 
These are the lines with smaller widths and most symmetric profiles. 
Wider and asymmetric lines are likely generated by line blends of multiple atomic species or molecules. 
We searched  the identified lines in the NIST atomic database and
we found counterparts for 22 of the 39 missing lines, but
only 7 had all atomic parameters measured. 
We added these lines to the atomic list used by the spectral synthesis code, 
but obtained no improvement in the spectrum produced. This means that the missing lines
likely have major contributions of other species.

Because some lines found in NIST lacked the atomic parameters needed to include them in the line
list, we also looked for them on VALD. We found counterparts for 14 of these 22 lines,
although 8 of them were already included in the line list adopted here.
For some of these lines we found more than one counterpart for a given line. All the lines from VALD 
not present in the atomic line list were from \ion{Fe}{i} and \ion{Fe}{ii}. Since the atomic parameters from VALD are
mostly not produced by measurements in laboratory, but are instead determined empirically or theoretically, 
we attempted to calibrate the
atomic parameters of these lines before including them in the line list. 

We used the ALiCCE code \citep{M14} to calibrate the oscillator strength of the Fe lines missing in the line list.
We did not try to calibrate the broadening parameters since the width of the lines is dominated 
by the rotational velocity of the Sun. The ALiCCE code found better log {\it gf} values for 5 of the 10 lines. 
The new values significantly improved the reproduction of the solar spectrum.
For the remaining lines the code has not been able to find results that could improve the reproduction 
of the solar spectrum.

        \begin{acknowledgements}
We would like to thank Robert Kurucz, the referee of this paper, for his valuable comments 
that definitely improved the paper. 
We would like to thank Gillian Nave for helping with the search of some of the unidentified lines.
J.R.K. acknowledges FAPESP (2014/00502-9) and CAPES for financial support.
L.M. thanks CNPQ for financial support through grant 303697/2015-6 and FAPESP through
grant 2015/14575-0.
P.C. acknowledges CNPQ for financial support through grant 305066/2015-3.
        \end{acknowledgements}

\bibliography{references} 
\bibliographystyle{aa}      

\end{document}